\documentclass[conference]{IEEEtran}
\IEEEoverridecommandlockouts
\IEEEpubid{\makebox[\columnwidth]{%
\parbox{\columnwidth}{\footnotesize
\copyright~2026 IEEE. Personal use of this material is permitted.
Permission from IEEE must be obtained for all other uses, in any current
or future media, including reprinting/republishing this material for
advertising or promotional purposes, creating new collective works, for
resale or redistribution to servers or lists, or reuse of any copyrighted
component of this work in other works.}%
\hfill} \hspace{\columnsep}\makebox[\columnwidth]{}}

\usepackage{cite}
\usepackage{amsmath,amssymb,amsfonts}
\usepackage{algorithmic}
\usepackage{graphicx}
\usepackage{textcomp}
\usepackage{xcolor}
\def\BibTeX{{\rm B\kern-.05em{\sc i\kern-.025em b}\kern-.08em
    T\kern-.1667em\lower.7ex\hbox{E}\kern-.125emX}}
\begin{document}

\title{Energy-Aware Performance Evaluation of Nonlinear Mechatronic Systems Under Matched-Tracking Conditions\\

}

\author{\IEEEauthorblockN{Sachintha Bhanuka Dayawansa}
\IEEEauthorblockA{\textit{Department of Electronic and Telecommunication Engineering} \\
\textit{University of Moratuwa}\\
Katubedda, Sri Lanka \\
dayawansasb.22@uom.lk}

}


\maketitle
\IEEEpubidadjcol

\begin{abstract}
Trajectory-tracking metrics such as root-mean-square error (RMSE), overshoot, and settling time are widely used to evaluate control performance in mechatronic systems. However, these measures describe output tracking alone and do not account for the actuator effort required to produce the observed motion. This limitation becomes more pronounced in nonlinear systems, where stiffness and dissipation depend on the system state.
This paper examines how these effects influence actuator energy under similar tracking conditions. An energy-aware evaluation framework is introduced that combines tracking error with cumulative actuator energy and enables comparison between systems with approximately matched performance. A simple index (EAPI) is used to capture both aspects in a single measure.
Simulation results for linear and nonlinear systems under proportional–derivative control show that comparable tracking accuracy can correspond to significantly different actuator energy. The nonlinear system consistently requires more energy across matched operating points, reflecting the influence of nonlinear stiffness and friction.
These results suggest that trajectory-based metrics alone may not fully capture differences in system effort, and that including energy provides a more informative basis for performance evaluation.
\end{abstract}


\section{Introduction}
Trajectory-tracking performance is a standard criterion used to  evaluate mechatronic systems, including robotic manipulators, actuators, and autonomous platforms. Conventional assessment relies on metrics such as rise time, settling time, overshoot, and root-mean-square error (RMSE), which quantify how closely system outputs follow a desired reference \cite{b1,b2}. These measures are widely used because they provide a clear and practical description of tracking behavior\cite{b3}.

However, prior work has noted that trajectory-based metrics capture only the similarity in generated trajectories and do not explicitly account for the physical effort required to generate that behavior \cite{b4,b5,b6}. This limitation becomes more prominent in nonlinear systems, where energy storage, dissipation, and actuator constraints depend on state and input in a nontrivial, highly coupled manner \cite{b7,b8}. As a result, systems with similar tracking responses may differ significantly in actuator energy consumption \cite{b9}.

Energy-related considerations have been explored in optimal control and energy-aware design, particularly in robotics and electromechanical systems \cite{b10,b11,b12,b13}. These studies often focus on minimizing control effort or improving efficiency under specific formulations. However, the evaluation problem itself has received limited attention, particularly regarding whether commonly used tracking metrics can adequately capture and distinguish differences between systems in terms of energy. 

This paper examines this question for nonlinear mechatronic systems, demonstrating that comparable tracking performance does not guarantee comparable actuator energy usage, especially in the presence of nonlinear stiffness and dissipative effects. These nonlinearities alter how energy is stored and dissipated, leading to differences that are not reflected in trajectory-based measures.

To study this effect an energy-aware evaluation framework is introduced that combines tracking metrics with a measure of cumulative actuator energy. To ensure meaningful comparison, systems are evaluated under matched, identical tracking conditions, so that differences in energy usage are not simply due to differences in tracking accuracy. Simulation results demonstrate that nonlinear systems can require substantially more actuator energy than linear counterparts, even when tracking performance is nearly identical. While it is understood that nonlinearities influence energy consumption, this work demonstrates that such energetic differences remain systematically undetectable under standard trajectory-based evaluation. The proposed matched-performance framework provides a structured method to expose this discrepancy. Unlike energy-optimal control methods that modify the controller to reduce effort, this work addresses the evaluation problem: whether two systems with similar tracking accuracy should be judged equivalent when their actuator-energy requirements differ. The contribution is therefore not a new controller, but a matched-performance assessment procedure that separates tracking quality from energetic burden. The proposed EAPI is used as a compact reporting index, while the main contribution is the matched-error comparison framework that exposes energy differences hidden by trajectory-only metrics.


\section{Energy-Aware Nonlinear Framework}

Consider the nonlinear mechatronic system
\begin{equation}
m\ddot{x} + \mathcal{D}(\dot{x}) + \mathcal{R}(x) = u,
\end{equation}
where $x(t) \in \mathbb{R}$ is the generalized coordinate, $m > 0$ is inertia, and $u(t)$ is the generalized actuator force/torque applied along coordinate x(t).    

To capture nonlinear energetic effects, the dissipative and restoring components are modeled as
\begin{equation}
\mathcal{D}(\dot{x}) = c\dot{x} + F_c\,\mathrm{sgn}(\dot{x}), \quad
\mathcal{R}(x) = kx + \alpha x^3,
\end{equation}
where $c > 0$ is viscous damping, $F_c > 0$ represents Coulomb friction, and $\alpha > 0$ introduces nonlinear stiffness.

The resulting dynamics become
\begin{equation}
m\ddot{x} + c\dot{x} + F_c\,\mathrm{sgn}(\dot{x}) + kx + \alpha x^3 = u.
\end{equation}

This formulation retains both amplitude-dependent stiffness and nonlinear dissipation, which directly influence energy storage and loss during system evolution.

\subsection{Mechanical Energy Representation}

The total stored mechanical energy can be defined as
\begin{equation}
H(x, \dot{x}) = T(\dot{x}) + V(x),
\end{equation}
where the kinetic energy is given by
\begin{equation}
T(\dot{x}) = \frac{1}{2} m \dot{x}^2,
\end{equation}
and the potential energy satisfies
\begin{equation}
\frac{dV}{dx} = \mathcal{R}(x).
\end{equation}

Integrating the restoring force yields
\begin{equation}
V(x) = \int_0^x (k s + \alpha s^3)\, ds = \frac{1}{2} k x^2 + \frac{1}{4} \alpha x^4.
\end{equation}

Hence, the total energy becomes
\begin{equation}
H(x, \dot{x}) = \frac{1}{2} m \dot{x}^2 + \frac{1}{2} k x^2 + \frac{1}{4} \alpha x^4.
\end{equation}

The additional quartic term introduces nonlinear energy growth that is not present in purely linear systems.

\subsection{Energy Balance}

Differentiating $H$ with respect to time gives
\begin{equation}
\dot{H} = m\dot{x}\ddot{x} + kx\dot{x} + \alpha x^3 \dot{x}.
\end{equation}

Substituting the system dynamics results in
\begin{equation}
\dot{H} = \dot{x}(u - c\dot{x} - F_c\,\mathrm{sgn}(\dot{x})),
\end{equation}
which simplifies to
\begin{equation}
\dot{H} = u\dot{x} - c\dot{x}^2 - F_c |\dot{x}|.
\end{equation}

This expression represents the energy balance of the system, where actuator input contributes power $u\dot{x}$, while viscous damping and Coulomb friction dissipate energy.

\subsection{Actuator Energy}

The instantaneous actuator power is defined as
\begin{equation}
P(t) = u(t)\dot{x}(t).
\end{equation}

The cumulative actuator energy over a time interval $[0, T]$ is given by
\begin{equation}
E(T) = \int_0^T \left| u(t)\dot{x}(t) \right| dt.
\end{equation}

This quantity measures the total energy transferred by the actuator, independent of direction, and depends on both control input and system response. The use of absolute power reflects total actuator effort rather than net energy flow. This is appropriate for systems with limited energy recovery or where actuator loading and thermal effects matter. While alternative methods such as signed energy can be used for regenerative systems, it does not reflect cumulative effort in the same way. However, in many practical mechatronic systems, including electric, hydraulic, and friction-dominated actuators, bidirectional power flow does not imply recoverable energy, as both positive and negative work contribute to losses through heat and internal dissipation. Therefore, integrating absolute power provides a more meaningful measure of actuator effort and energy burden than signed energy.

\subsection{Tracking Metrics and Energetic Limitation}

Let $x_d(t)$ denote the desired trajectory and define the tracking error
\begin{equation}
e(t) = x_d(t) - x(t).
\end{equation}

A commonly used performance measure is the root-mean-square error
\begin{equation}
\mathrm{RMSE} = \sqrt{\frac{1}{T} \int_0^T e^2(t)\, dt}.
\end{equation}

However, this metric depends only on trajectory deviation and does not include actuator input, velocity, or energy-related quantities. As a result, two systems can exhibit similar RMSE values while requiring different levels of actuator energy.

\subsection{Combined Performance Measure}

To evaluate both tracking accuracy and energetic cost, define
\begin{equation}
J_e = \int_0^T e^2(t)\, dt, \quad
J_u = \int_0^T \left| u(t)\dot{x}(t) \right| dt.
\end{equation}

Using reference values $J_e^{\mathrm{ref}}$ and $J_u^{\mathrm{ref}}$, a normalized performance index is introduced as
\begin{equation}
\mathrm{EAPI} = \lambda \frac{J_e}{J_e^{\mathrm{ref}}} + (1 - \lambda)\frac{J_u}{J_u^{\mathrm{ref}}}, \quad 0 \leq \lambda \leq 1.
\end{equation}

Here, $J_e^{\mathrm{ref}}$ and $J_u^{\mathrm{ref}}$ are reference normalization constants used to scale the tracking and energy terms to comparable magnitudes. In this study, they are chosen from the baseline linear operating condition used for normalization. The weighting factor $\lambda = 0.5$ is used to assign equal importance to tracking error and actuator energy; other choices of $\lambda$ reflect application-specific priorities. This formulation enables joint evaluation of tracking performance and actuator energy, allowing differences in energetic behavior to be captured even when trajectory-based metrics appear similar.

\section{Energetic Differences Under Nonlinear Dynamics}

This section examines how nonlinear dynamics influence actuator energy requirements relative to linear systems. The analysis shows that differences in energy storage and dissipation can lead to variations in actuator energy, even when trajectory-tracking performance is similar.

\subsection{Linear and Nonlinear Energy Structure}

Consider the nonlinear system
\begin{equation}
m\ddot{x} + c\dot{x} + F_c\,\mathrm{sgn}(\dot{x}) + kx + \alpha x^3 = u_N,
\end{equation}
and the corresponding linear system
\begin{equation}
m\ddot{x} + c\dot{x} + kx = u_L.
\end{equation}

The associated stored mechanical energies are
\begin{equation}
H_N = \frac{1}{2} m \dot{x}^2 + \frac{1}{2} k x^2 + \frac{1}{4} \alpha x^4,
\end{equation}
\begin{equation}
H_L = \frac{1}{2} m \dot{x}^2 + \frac{1}{2} k x^2.
\end{equation}

The difference in stored energy is therefore
\begin{equation}
\Delta H(x) = H_N - H_L = \frac{1}{4} \alpha x^4.
\end{equation}

Since $\alpha > 0$, the nonlinear system includes an additional amplitude-dependent energy component. This term increases with displacement magnitude and contributes to higher stored energy levels compared to the linear case.

\subsection{Dissipative Effects}

From the energy balance derived in Section II, the nonlinear system satisfies
\begin{equation}
\dot{H}_N = u_N \dot{x} - c\dot{x}^2 - F_c |\dot{x}|,
\end{equation}
while the linear system satisfies
\begin{equation}
\dot{H}_L = u_L \dot{x} - c\dot{x}^2.
\end{equation}

The cumulative dissipated energy over the interval $[0, T]$ is therefore
\begin{equation}
D_N(T) = \int_0^T \left( c\dot{x}^2 + F_c |\dot{x}| \right) dt,
\end{equation}
\begin{equation}
D_L(T) = \int_0^T c\dot{x}^2 \, dt.
\end{equation}

The difference becomes
\begin{equation}
D_N(T) - D_L(T) = F_c \int_0^T |\dot{x}| \, dt.
\end{equation}

For $F_c > 0$, this term is non-negative and increases with motion duration and velocity magnitude. This indicates that the nonlinear system dissipates additional energy through friction, which must be compensated by the actuator.

\subsection{Implications for Actuator Energy}

Integrating the energy balance over $[0, T]$ for both systems gives
\begin{equation}
\int_0^T u_N \dot{x} \, dt = H_N(T) - H_N(0) + D_N(T),
\end{equation}
\begin{equation}
\int_0^T u_L \dot{x} \, dt = H_L(T) - H_L(0) + D_L(T).
\end{equation}

Subtracting the two expressions yields
\begin{equation}
\int_0^T u_N \dot{x} \, dt - \int_0^T u_L \dot{x} \, dt 
= \left( H_N(T) - H_L(T) \right) + \left( D_N(T) - D_L(T) \right).
\end{equation}

Using the expressions derived above, this becomes
\begin{equation}
\int_0^T u_N \dot{x} \, dt - \int_0^T u_L \dot{x} \, dt 
= \frac{1}{4} \alpha x^4(T) + F_c \int_0^T |\dot{x}| \, dt.
\end{equation}

Both terms on the right-hand side are non-negative for $\alpha > 0$ and $F_c > 0$.

This shows that differences in nonlinear stiffness and friction introduce additional contributions to actuator work that are not present in the linear system.

\subsection{Interpretation Under Comparable Tracking}

This shows that even under similar tracking accuracy, actuator energy can differ significantly because nonlinear stiffness adds amplitude-dependent energy storage ($\alpha x^4$) and Coulomb friction introduces additional dissipation ($F_c |\dot{x}|$); therefore, actuator effort depends on both trajectory evolution and internal energy mechanisms, meaning trajectory-based metrics alone cannot fully capture system performance differences.

\section{Matched-Performance Evaluation Framework}

Direct comparison of actuator energy between dynamical systems can be misleading if the systems achieve different levels of tracking accuracy. A system that tracks more accurately may naturally require greater control effort, making it difficult to attribute energy differences solely to the underlying dynamics. To ensure a meaningful comparison, actuator energy is evaluated under approximately equivalent tracking conditions.

\subsection{Performance-Constrained Comparison}

For a given system $\Sigma_i$, we define the tracking and energetic measures
\begin{equation}
J_{e,i} = \int_0^T e_i^2(t)\, dt, \quad
E_i(T) = \int_0^T \left| u_i(t)\dot{x}_i(t) \right| dt.
\end{equation}

A comparison between two systems is considered valid only when their tracking performance is sufficiently close, i.e.,
\begin{equation}
\left| J_{e,1} - J_{e,2} \right| \leq \epsilon_m,
\end{equation}
where $\epsilon_m > 0$ is a prescribed tolerance.

This condition ensures that differences in actuator energy are not primarily caused by differences in tracking quality.

\subsection{Parameterized Controller Exploration}

To generate comparable operating conditions, a PD controller is used:
\begin{equation}
u(t) = K_p e(t) + K_d \dot{e}(t),
\end{equation}
with parameters
\begin{equation}
\theta = (K_p, K_d).
\end{equation}

For each parameter choice $\theta$, simulation produces a pair
\begin{equation}
\Phi(\theta) = \left( J_e(\theta), E(\theta) \right),
\end{equation}
which maps controller parameters to tracking-error and energy values.

Applying this procedure separately to the linear and nonlinear systems yields two sets of achievable operating points
\begin{equation}
\Phi_L(\theta) = (J_{e,L}, E_L), \quad
\Phi_N(\theta) = (J_{e,N}, E_N).
\end{equation}

These sets describe how tracking performance and actuator energy vary across different controller configurations.

\subsection{Matched-Performance Pairing}

For a given operating point of the linear system $\theta_L$, a corresponding nonlinear operating point $\theta_N$ is selected such that the tracking error difference is minimized, subject to the matching condition.

Specifically, $\theta_N$ is chosen as
\begin{equation}
\theta_N^* = \arg \min_{\theta_N} \left| J_{e,N}(\theta_N) - J_{e,L}(\theta_L) \right|,
\end{equation}
subject to
\begin{equation}
\left| J_{e,N}(\theta_N) - J_{e,L}(\theta_L) \right| \leq \epsilon_m.
\end{equation}

This procedure ensures that each comparison is performed between operating points with similar tracking accuracy.

\subsection{Energetic Comparison}

Once a matched pair $(\theta_L, \theta_N)$ is obtained, the difference in actuator energy is evaluated as
\begin{equation}
\Delta E = E_N(\theta_N) - E_L(\theta_L).
\end{equation}

To express this difference in relative terms, the percentage increase is computed as
\begin{equation}
\Gamma_E = \frac{E_N - E_L}{E_L} \times 100\%.
\end{equation}

This metric quantifies how much additional actuator energy is required by the nonlinear system under comparable tracking conditions.

\subsection{Interpretation}

In practice, the matching process is carried out by sweeping controller gains over a predefined parameter space and computing the corresponding $(J_e, E)$ values for each system. For each linear operating point, the closest nonlinear operating point in terms of tracking error is identified within the specified tolerance. By sweeping controller gains to generate matched-performance pairs with similar tracking error, the framework enables consistent comparison of actuator energy across operating conditions, isolating the effect of system dynamics so that observed energy differences can be attributed to nonlinear storage and dissipation mechanisms rather than control accuracy. 
\section{Results and Discussion}

To validate the proposed energy-aware evaluation framework, simulation studies were conducted for both the linear and nonlinear mechatronic systems under similar tracking conditions using proportional-derivative (PD) control. The simulations were designed to investigate whether comparable trajectory-tracking performance necessarily implies comparable energetic behavior. Simulations used MATLAB \texttt{ode45} with $T=8~\mathrm{s}$, $m=1.0$, $c=0.35$, $k=25$, $\alpha=300$, $F_c=0.8$, $K_p\in[40,180]$, $K_d\in[6,30]$, and matching tolerance $\epsilon_m=0.003$.

\subsection{Cumulative Actuator Energy Evolution}

\begin{figure}[!t]
    \centering
    \includegraphics[width=\linewidth]{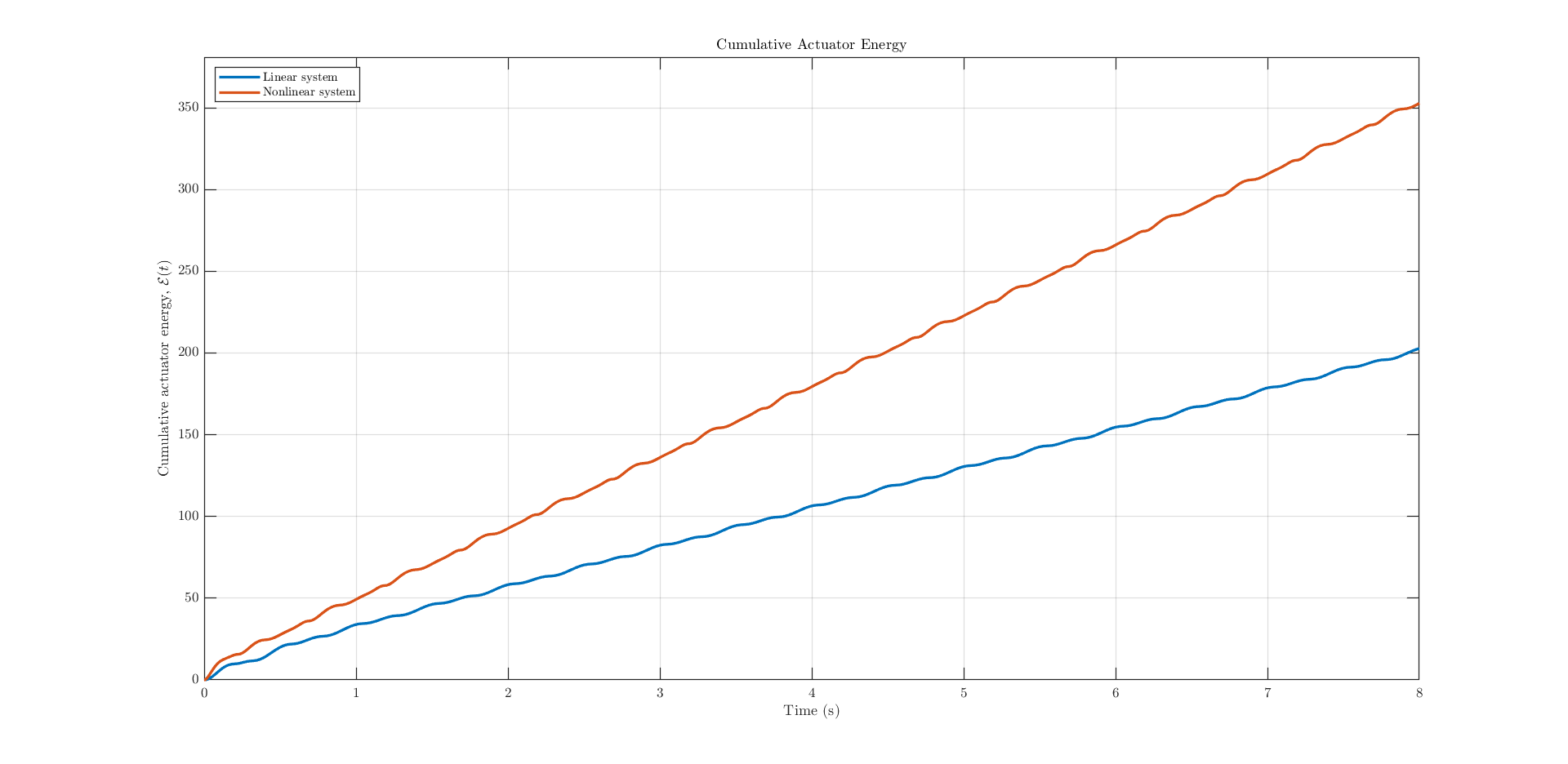}
    \caption{Cumulative actuator energy comparison between linear and nonlinear systems under trajectory tracking.}
    \label{fig:cumulative_energy}
\end{figure}

The cumulative actuator energy evolution for the linear and nonlinear systems is shown in Fig.~\ref{fig:cumulative_energy}. The cumulative energy was computed using the proposed actuator-energy formulation

\begin{equation}
\mathcal{E}(T) = \int_{0}^{T} \left| u(t)\dot{x}(t) \right| dt
\end{equation}

which represents the total actuator energy expenditure during system execution.

The nonlinear system accumulates energy more rapidly over time. By the end of the simulation, it reaches roughly 350 units, compared to about 200 units for the linear case. This difference is consistent with the model structure: the nonlinear system includes additional stiffness and friction terms, which increase both stored energy and dissipation. As a result, more control effort is required even when the output trajectories are similar. These results therefore demonstrate that trajectory similarity alone cannot characterize the underlying energetic behavior of nonlinear systems.

\subsection{Phase-Space Energetic Distortion}

\begin{figure}[!t]
    \centering
    \includegraphics[width=\linewidth]{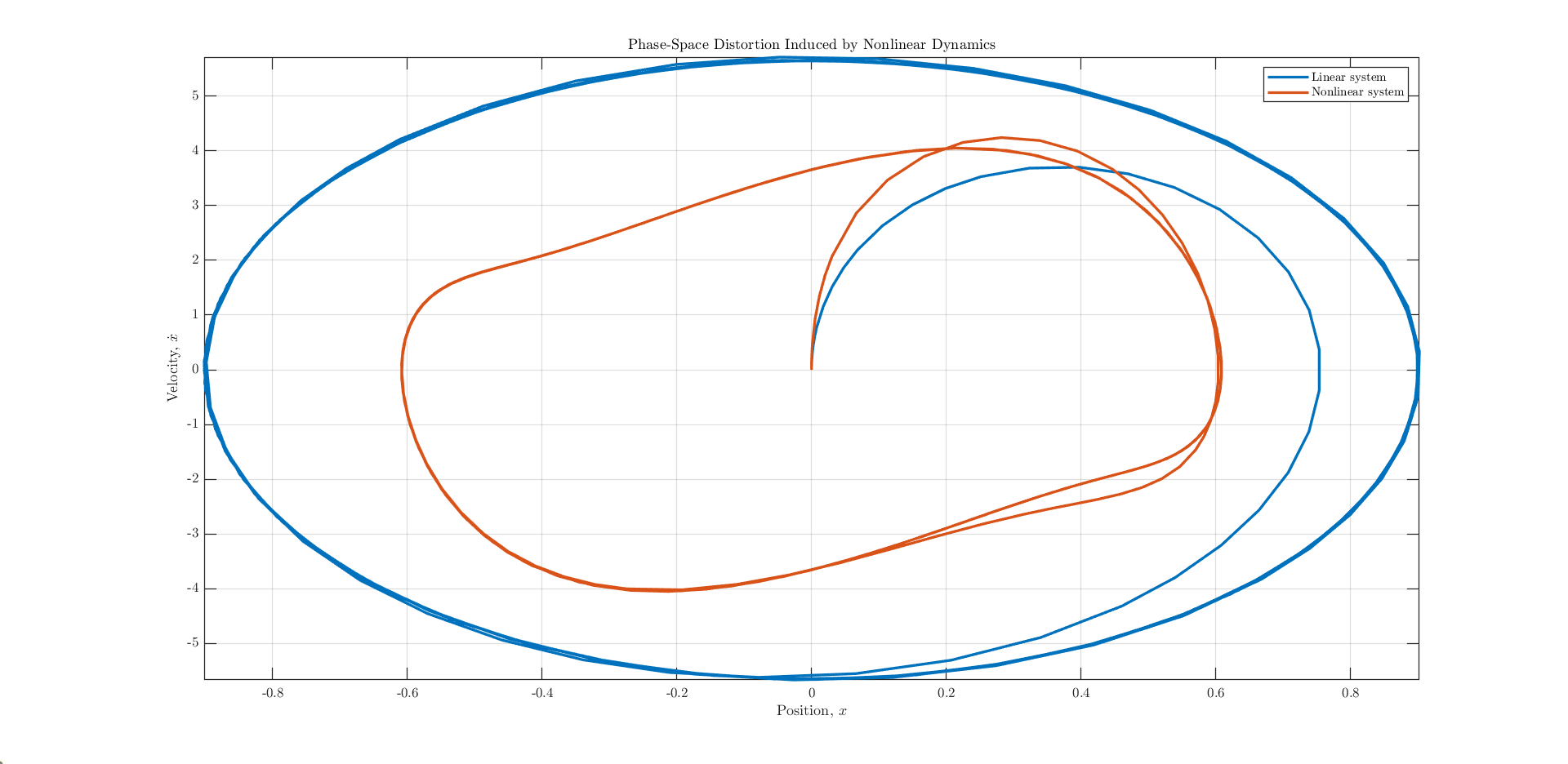}
    \caption{Phase-space trajectories illustrating energetic distortion induced by nonlinear dynamics.}
    \label{fig:phase_space}
\end{figure}

To investigate how nonlinear dynamics alter the energetic geometry of the system, the phase-space trajectories of the linear and nonlinear systems were examined. The resulting phase portraits are shown in Fig.~\ref{fig:phase_space}.

The linear system exhibits approximately elliptical phase-space trajectories, consistent with the classical quadratic energy structure associated with linear oscillatory systems. In contrast, the nonlinear system produces substantially distorted trajectories characterized by asymmetric and amplitude-dependent deformation. This result experimentally validates the theoretical analysis presented in Section III-C, where it was shown that the nonlinear energy function modifies the geometry of the energy manifolds through the quartic energy term. The nonlinear system therefore evolves along fundamentally different energetic trajectories despite maintaining comparable motion behavior in the output space.

\subsection{Matched-Performance Energy Comparison}

To ensure that energetic comparison remained scientifically meaningful, matched-performance evaluation was performed according to the framework introduced in Section IV. Linear and nonlinear operating points were paired such that the corresponding RMSE values remained approximately equal within the prescribed matching tolerance. This procedure ensured that energetic differences originated primarily from the intrinsic nonlinear dynamics rather than differences in tracking quality.

\begin{figure}[!t]
    \centering
    \includegraphics[width=\linewidth]{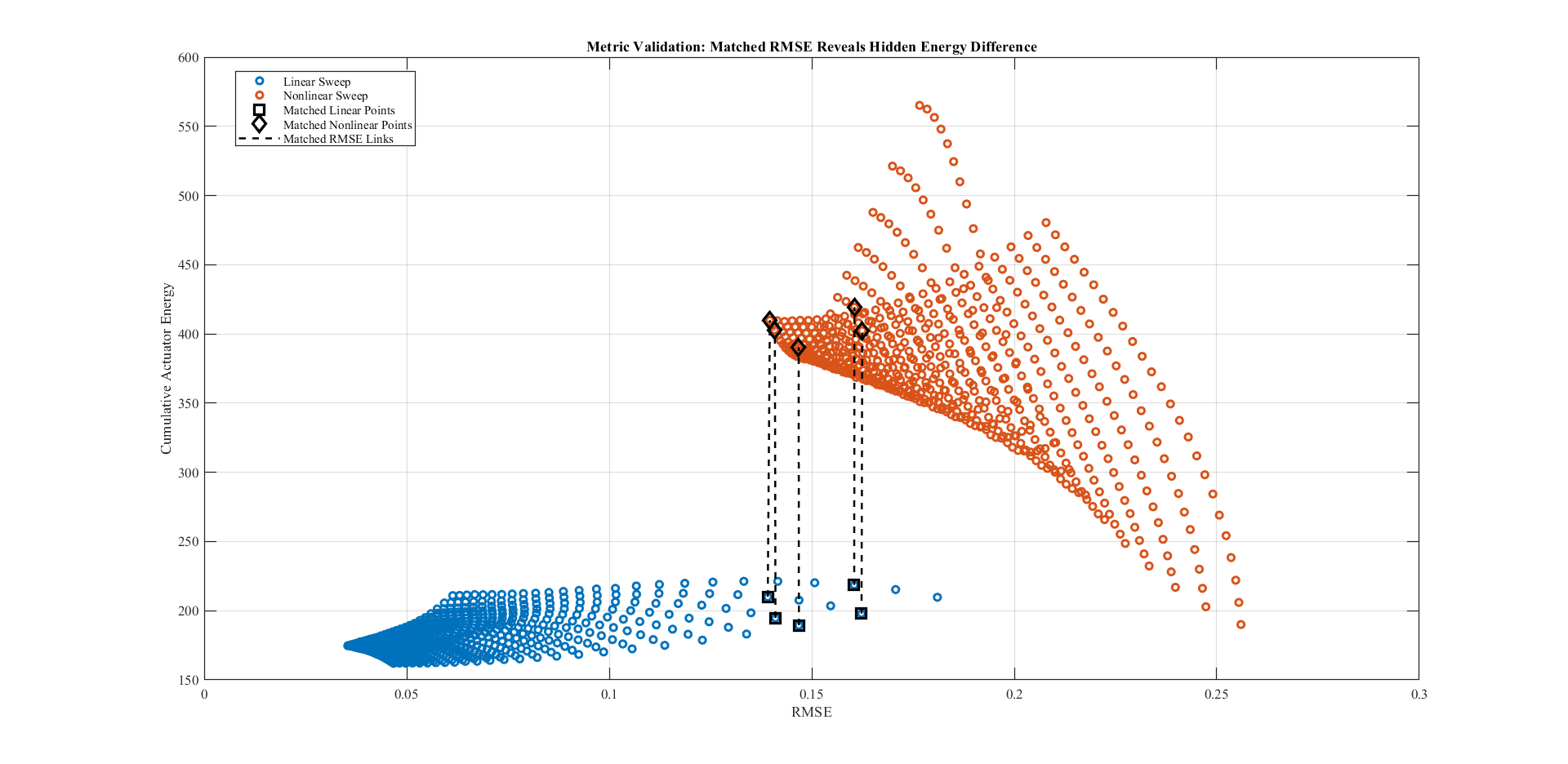}
    \caption{RMSE-energy operating manifolds for linear and nonlinear systems with matched-performance pairing links.}
    \label{fig:rmse_energy}
\end{figure}

The resulting matched operating points are illustrated in Fig.~\ref{fig:rmse_energy}, which shows the relationship between RMSE and cumulative actuator energy for both systems across extensive controller-gain sweeps.

The strongest matched-performance result obtained during the simulations is summarized below:

\begin{itemize}
    \item Linear system: RMSE $= 0.14104$, Energy $= 194.727$
    \item Nonlinear system: RMSE $= 0.14089$, Energy $= 402.690$
\end{itemize}

This corresponds to an increase of approximately $106.8\%$ in actuator energy despite nearly identical tracking error.
Hence, this result is consistent with the theoretical analysis established in Section III-E:

\begin{equation}
\mathcal{M}(\Sigma_1) \approx \mathcal{M}(\Sigma_2) \Rightarrow \mathcal{E}_1 \not\approx \mathcal{E}_2
\end{equation}

\subsection{Matched-Performance Comparison}

\begin{figure}[!t]
    \centering
    \includegraphics[width=\linewidth]{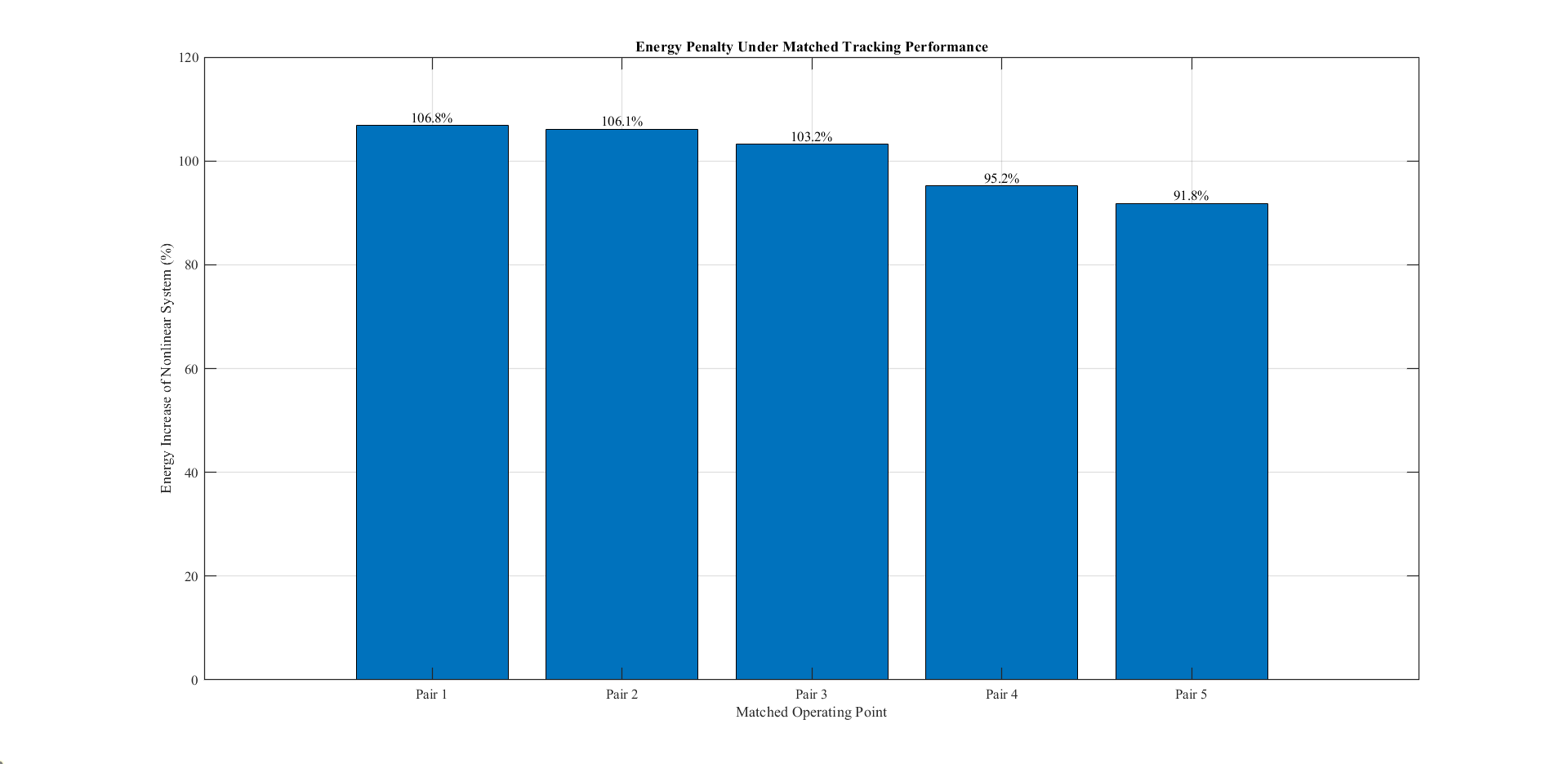}
    \caption{RMSE vs. actuator energy for linear and nonlinear systems with matched pairs.}
    \label{fig:energy_divergence}
\end{figure}

The relative energy increase ranges from $\sim 92\%$ to $\sim 107\%$ across matched-performance pairs, showing that the nonlinear system consistently requires nearly twice the actuator energy of the linear system. Since this trend appears across multiple matched pairs, the difference is attributable to the nonlinear system structure rather than a specific controller choice. This has practical implications for battery life, thermal loading, and component wear, supporting the need for energy-aware evaluation in addition to trajectory-tracking metrics.

\subsection{Energy-Aware Performance Index (EAPI) }

\begin{figure}[!t]
    \centering
    \includegraphics[width=\linewidth]{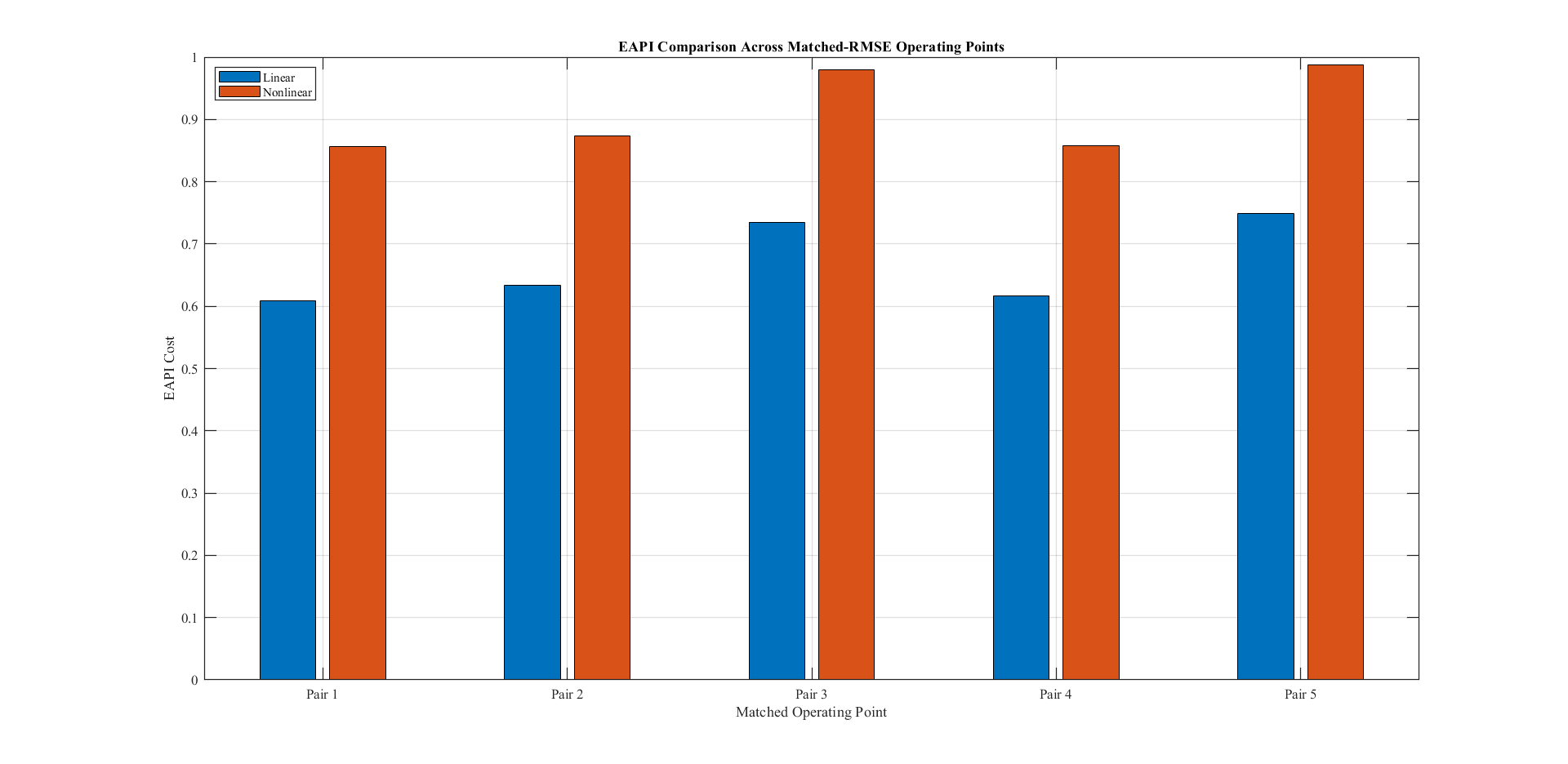}
    \caption{Comparison of Energy-Aware Performance Index (EAPI) values for matched linear and nonlinear operating points.}
    \label{fig:eapi_comparison}
\end{figure}

To jointly evaluate tracking accuracy and energy usage, the proposed Energy-Aware Performance Index (EAPI) was computed for all matched operating pairs. The resulting EAPI comparison is shown in Fig.~\ref{fig:eapi_comparison}.

The nonlinear system consistently produced larger EAPI values than the linear system across all matched operating conditions, reflecting higher energy cost for similar tracking. For the strongest matched-performance case, the linear system achieved an EAPI value of $0.60905$, whereas the nonlinear system produced an EAPI value of $0.85627$, corresponding to an increase of approximately $40.59\%$.

This result demonstrates an important limitation of conventional trajectory-based evaluation. Because RMSE values remain nearly identical for the matched systems, a purely trajectory-based assessment would incorrectly suggest comparable performance. However, the EAPI formulation successfully reveals the hidden energetic inefficiency associated with the nonlinear dynamics. The proposed energy-aware metric therefore provides significantly improved discriminatory capability compared with classical trajectory-only evaluation methods.

\begin{table}[!t]
\caption{Matched RMSE and actuator-energy comparison between linear and nonlinear systems}
\label{tab:energy_comparison}
\centering
\resizebox{\columnwidth}{!}{%
\begin{tabular}{|c|c|c|c|c|c|}
\hline
Pair & Linear RMSE & Nonlinear RMSE & Linear Energy & Nonlinear Energy & Energy Increase (\%) \\
\hline
Pair 1 & 0.1410 & 0.14089 & 194.73 & 402.69 & 106.80 \\
\hline
Pair 2 & 0.1468 & 0.14676 & 189.44 & 390.52 & 106.14 \\
\hline
Pair 3 & 0.1623 & 0.16246 & 197.98 & 402.33 & 103.22 \\
\hline
Pair 4 & 0.1391 & 0.13965 & 209.98 & 409.87 & 95.19 \\
\hline
Pair 5 & 0.1605 & 0.16053 & 218.63 & 419.29 & 91.78 \\
\hline
\end{tabular}%
}
\end{table}

Tables I and II show that this trend holds across all tested pairs, with consistent increases in both energy and EAPI.

\begin{table}[!t]
\caption{EAPI comparison under matched-performance operating conditions}
\label{tab:eapi_comparison}
\centering
\begin{tabular}{|c|c|c|c|}
\hline
Pair & EAPI (Linear) & EAPI (Nonlinear) & EAPI Increase (\%) \\
\hline
Pair 1 & 0.60905 & 0.85627 & 40.593 \\
\hline
Pair 2 & 0.63426 & 0.87374 & 37.758 \\
\hline
Pair 3 & 0.73505 & 0.97978 & 33.294 \\
\hline
Pair 4 & 0.61717 & 0.85821 & 39.055 \\
\hline
Pair 5 & 0.74862 & 0.98817 & 31.999 \\
\hline
\end{tabular}
\end{table}

Overall, the simulation results consistently demonstrate that nonlinear systems can require significantly more actuator energy even when tracking performance is similar. This suggests that trajectory-based metrics alone may not fully reflect differences in system effort, while including energy provides a more complete comparison.

\section{Conclusion}
This paper examined the limits of trajectory-based evaluation for nonlinear mechatronic systems, showing that similar tracking performance does not necessarily correspond to similar actuator energy. An energy-aware framework was introduced that combines tracking error with actuator energy and enables comparison under matched tracking conditions.

The analysis shows that nonlinear stiffness and friction can increase energy requirements relative to linear systems, even when tracking accuracy is similar. Simulation results support this observation, with the nonlinear system requiring substantially more actuator energy under closely matched RMSE conditions . The proposed index (EAPI) captures these differences by incorporating both tracking and energy into a single measure.

Overall, the results suggest that including energy alongside trajectory metrics provides a more informative basis for comparing system performance. Future work will consider extension to higher-dimensional systems and experimental validation.

\end{document}